1

# Unquenching the Schwinger model * †

Alan C. Irving[a] and James C. Sexton[b]

[a]DAMTP, University of Liverpool
PO Box 147, Liverpool L69 3BX, UK

[b]School of Mathematics, Trinity College,
Dublin 2, Ireland

We study the quenched and unquenched lattice Schwinger model with Wilson fermions. The lowest non-trivial order of the systematic expansion recently proposed by Sexton and Weingarten is shown to allow good estimates of long distance physics from quenched configurations. Results for the static potential and the lowest bound state mass are presented.

## 1. INTRODUCTION

Renewed efforts have recently been made to design improved algorithms for the simulation of theories with dynamical fermions. The pressing nature of the problem in QCD is well known. Since much of the effect of dynamical quarks in QCD is predicted by perturbative arguments to amount to a shift in the effective lattice spacing, it will require considerable computational power and improved algorithms to isolate meaningful and non-trivial effects for quantities of physical interest. The Schwinger model offers a suitable testing ground of these questions for several reasons. The quenched lattice theory has (unlike QCD) an almost trivial pure gauge sector with most of the dynamics contributed by the fermions. The effects of unquenching are expected to be pronounced. Much is known exactly and perturbatively about the corresponding continuum theory. Obviously the Schwinger lattice model is computationally more accessible than full QCD.

Sexton and Weingarten have recently proposed a systematic expansion of the fermion determinant in terms of its constituent gauge invariant loops [1],[2]. An initial test with QCD on relatively small systems ($6^4$) has given promising results in comparison with hybrid Monte Carlo simulation of the full theory. They showed [2] that the discrepancy between Wilson loops measured in quenched and full simulations can be accounted for by the lowest non-trivial systematic correction estimate given by this expansion. This might seem surprising at first sight. The systematic correction, or deficit, is estimated at lowest order from an approximate evaluation of the overlaps of an observable with the trace log of the fermion matrix and with the smallest Wilson loop (plaquette) – both with respect to the quenched vacuum. One might expect this estimate to be severely tested when truly long range physics is studied. With this in mind, we have begun a study of the Schwinger model. We wish to test the limits of the method and to uncover any physical insights relevant to the Schwinger model itself and to the continuing work on QCD.

## 2. METHOD

Details of the method can be found in [1],[2]. The implementation for the Schwinger model is essentially the same as for QCD. We briefly review some of the basic ideas and formulae. We use the Wilson formulation of the lattice Schwinger action

$$S = S_G + S_f \quad \text{where} \quad S_f = \overline{\psi} M[U] \psi \qquad (1)$$

---

in a suppressed index notation. The fermion matrix M can be decomposed into its red/black (even/odd) components

$$M = 1 + \kappa K = \begin{pmatrix} 1 & \kappa K^{oe} \\ \kappa K^{eo} & 1 \end{pmatrix}. \qquad (2)$$

From this one may construct a Hermitian, positive definite, red/black preconditioned matrix

$$H^{(e)} = M^{(e)} M^{(e)\dagger}, \qquad M^{(e)} = 1 - \kappa^2 K^{eo} K^{oe} \qquad (3)$$

from which one may reconstruct the effective fermion-gauge action for $n_f$ flavours of fermion $S_{\text{eff}} = -n_f T/2$, where

$$T \equiv \text{Tr}\ln(MM^\dagger) = 2\ln\det M = \ln\det H^e. \qquad (4)$$

We may formally expand $T$ in terms of a Gram-Schmidt orthogonalised basis of ordered gauge-invariant loops, suitably symmetrised over spatial orientations and translations:

$$T = \sum_{t=0}^{\infty} = a_t \hat{S}_t = L_n + R_n \qquad (5)$$

where $L_n$ represents a truncation after $n$ terms and $R_n$ the corresponding remainder. Here $\hat{S}_0 = 1$, (loops of size 0) and

$$\hat{S}_1 = S_1 - <S_1> \qquad (6)$$

where $S_1 \equiv S_G$ is the (unnormalised) plaquette action. The higher loops $S_n$ used to form the orthogonal basis are also unnormalised sums (e.g. $S_2$ on a unit $L \times L$ lattice has value $2L^2$). The basis vectors themselves satisfy

$$<\hat{S}_i \hat{S}_j> = <\hat{S}_i^2> \delta_{ij}, \quad <\hat{S}_i> = 0 \; (i > 1). \qquad (7)$$

It is immediately obvious that the truncation $L_0$ is the quenched approximation and that $L_1$ corresponds to a shift in gauge coupling $\delta\beta$. Each truncation gives rise to an effective partition function

$$Z_n = \int [dU] \exp(-S_G + \frac{n_f}{2} L_n) \qquad (8)$$

and the corresponding expectation value $<A>_n$ of an arbitrary operator $A$. One can show [1],[2] that the deficit in such an estimate due to the neglect of $R_n$ is approximately

$$\Delta <A>_n = \frac{n_f}{2}(<AR_n>_n \\ - <A>_n <R_n>_n). \qquad (9)$$

In view of the above arguments, $<A>_1$ can be estimated from quenched configurations shifted in $\beta$. The corresponding deficit can then be obtained from

$$\Delta <A>_1 = \frac{n_f}{2}(<AT>_1 - <A>_1 <T>_1 \\ - a_1 <A\hat{S}_1>_1) \qquad (10)$$

where

$$a_1 = -\frac{2\delta\beta}{n_f} = \frac{<T\hat{S}_1>_1}{<\hat{S}_1^2>_1}. \qquad (11)$$

The stochastic evaluation of $T = \text{Tr}\ln M$ using Gaussian noise vectors and a Chebychev approximation to the logarithm is described in [1]. In the present work, we find 40 noise vectors are necessary to obtain fluctuations which are comparable with or smaller than those with respect to the gauge fluctuations.

An immediate test of the method is to compare $<A>$ from a full dynamical fermion simulation (FDFS) at $\beta_f = \beta_0 + \delta\beta$ with $<A>_1 + \Delta <A>_1$ estimated from quenched configurations at $\beta_0$ [1]. For this purpose, we take $A$ in turn to be Wilson loops, the static inter-fermion potential $V(r)$ derived from these, and $m_v$ the mass of the 'vector' bound state. The latter is known in the massless continuum model to be given by $e/\sqrt{\pi}$ where $e$ is the fermion charge. On the lattice we deduce it from the large Euclidean time dependence of zero momentum correlators involving $\overline{\psi}\gamma_1\psi$. These were constructed from 'quark' propagators which were obtained by a conjugate gradient-style solver applied to the red/black preconditioned matrix $H^{(e)}$. We tried further preconditioning in momentum space with a suitably tuned free fermion propagator. For the present application the acceleration achieved did not repay the overheads.

The FDFS was performed with the hybrid Monte Carlo algorithm [3] which is exact for even $n_f$. We choose $n_f = 2$ for the comparisons. The continuum version of the massive multi-flavour model is known to have a non-trivial spectrum on finite systems with light fermions [4]. In particular the limits $L \to \infty$ and $m \to 0$ do not commute. This will be a challenge for future detailed analyses of the flavour dependence of unquenched models.

## 3. RESULTS

Our preliminary analysis was performed on lattices of size $32^2$, at $\beta$ values from 2.0 to 3.0, and $\kappa$ from 0.25 to 0.265. For example, we generated 2000 quenched configurations at $\beta = 2.5$ which we first gauge-fixed (Landau) in case Fourier acceleration was required later. We used a Lanczos algorithm to estimate the maximum and minimum eigenvalues of $H^{(e)}$ at each $\kappa$ (needed for the Chebychev approximation) and then accumulated stochastic estimates of $T$ and $T^2$. We also accumulated measurements of the operators $A$ on each configuration. From these, we evaluated $\delta\beta$ for 2 flavours (Eqn. 11), some higher expansion coefficients $a_n$ (such as two and three plaquette operators) and deficit estimates $\Delta < A >_1$ for the above observables.

At $\beta_0 = 2.50$ and $\kappa = 0.26$, we found $\delta\beta = -0.212 \pm 0.010$. These and other all other errors were evaluated using bootstrap with subensembling to detect and remove autocorrelations. We then generated a further 2000 configurations incorporating dynamical fermions ($\kappa = 0.26$) at $\beta = \beta_0 + \delta\beta = 2.29$. A simple preliminary test is to compare $< S_1 >$ ($\equiv < S_G >$) for the quenched data, $783.3 \pm .3$ with the full fermion data, $782.8 \pm .3$ at shifted $\beta$. Since the above uncertainty in $\delta\beta$ translates to one of $\pm 1.4$ in $< S_G >$ this is a very satisfactory comparison.

Fig. 1 shows the vector state effective mass $m_v$ obtained from the FDFS at $\beta = 2.29$, $\kappa = 0.26$ in comparison with that estimated from the quenched data together with the deficit (10). The dashed line in this figure serves both to indicate the quenched value at $\beta = 2.29$, and the Euclidean time range used to extract mass estimates. From the FDFS and from the corrected estimate from the quenched data respectively we find $m_v = 0.369(3)$ and $0.367(7)$ where the bracketed digit is the error in the last significant digit. These are in complete agreement. In the corrected value, the contribution due to the $\beta$ shift is $-.041$ and that due to the truncation deficit, $-.037$.

Fig. 2 shows a comparison of the static potential using quenched data at $\beta = 2.50$, FDFS data at 2.29 and the corrected estimate from the quenched data. The results have been expressed in dimensionless units using our result $m_v = 0.37$ in lattice units. Also shown is the (exact) infinite volume quenched result (dashed line) and the continuum single flavour massless Schwinger result (full line) obtained by evaluating Wilson loops due to a vector field of mass $e/\sqrt{\pi}$. The continuum result is

$$\frac{V(m_v R)}{m_v} = \frac{\pi}{2}(1 - e^{-m_v R}). \qquad (12)$$

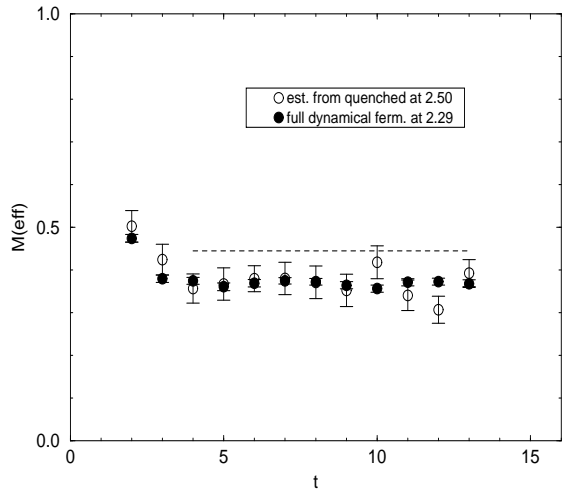

Figure 1. Effective masses from the FDFS and corrected quenched data at $\kappa = 0.26$.

The comparison is satisfactory although the errors associated with the deficit estimate (10) get quite large at large distances. There could be a systematic under-estimate at increasing $R$, but this is not yet statistically significant. The same effect is seen in a direct comparison using Wilson loops with increasing area (not shown here). However, it is clear that the potential screening effects due to light dynamical fermions are capable of being reproduced by these low order estimates.

In Table 1, we collect information on the relative contributions to the expansion of $T = 2\text{Tr}\ln M$. We display the coefficients $a_1, a_2$, $a_{3a}$ and $a_{3b}$ ($3a$ corresponds to three plaquettes in a row, $3b$ is a 'chair' of three plaquettes). Also listed is the corresponding root variance $\sigma_n$ of $\hat{S}_n$ and the product $a_n \sigma_n$. It is clear that the single plaquette provides the largest contribution to the fluctuations in $T$ and that the coefficients are decreasing. It is presumably for this reason that we can obtain reliable estimates of the remaining deficit incurred when working with quenched configurations at shifted $\beta$.

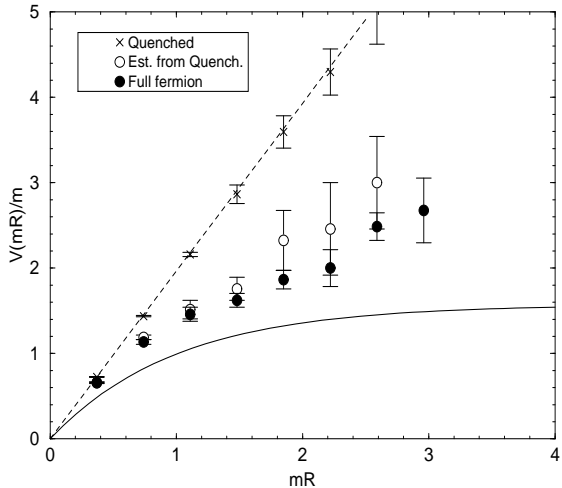

Figure 2. Static potential in physical units: quenched at $\beta = 2.50$ (crosses), corrected for the deficit at first order (open circles) and from FDFS (full circles). The curves are described in the text.

Table 1
Contributions to the fluctuations in $T = 2\mathrm{Tr}\ln M$.

|  | $a_n$ | $\sigma_n$ | Product |
|---|---|---|---|
| $\hat{S}_1$ | 0.212(10) | 10.7(2) | 2.27(11) |
| $\hat{S}_2$ | 0.085(7) | 15.2(2) | 1.29(11) |
| $\hat{S}_{3a}$ | 0.041(6) | 14.2(3) | 0.58(9) |
| $\hat{S}_{3b}$ | 0.025(4) | 25.2(4) | 0.63(10) |

## 4. COMMENTS AND CONCLUSIONS

The Schwinger model results reinforce those already obtained in QCD [1,2]. One may indeed use quenched configurations to study non-trivial physics due to dynamical fermions. However, the work required in the stochastic evaluation of $\mathrm{Tr}\ln M$ using current techniques is still unaccept- ably large. Having estimated the (small) higher coefficients in the expansion, one may now consider simulating with a higher order effective action. This could be done in the same spirit as above i.e. by making use of a deficit correction or, perhaps more promisingly, by using the effective action to generate candidate configurations within an exact simulation. These and other issues will be addressed in future work.